\documentclass[a4paper,12pt]{article}

\title{\Large Potential of a moving test charge in a dusty plasma in the presence of grain
size distribution and grain charging dynamics}

\author{\normalsize Michael A. Raadu\thanks{%
e-mail: michael.raadu@alfvenlab.kth.se} and \underline{Muhammad Shafiq}\thanks{%
e-mail: mohammad.shafiq@alfvenlab.kth.se} \\
\small Royal Institute of Technology, Alfv\'en Laboratory,\\ \small Division of Plasma Physics,
SE-100 44 Stockholm, Sweden}

\begin{document}
\maketitle

\begin{abstract}
It is well known that the form of grain size distribution strongly influences
the linear dielectric response of a dusty plasma. In previous results
[IEEE Trans. Plasma Sci. 29, 182 (2001)], it was shown that for a class of size
distributions, there is an equivalence to a Lorentzian distribution of mono-sized
particles. The electrostatic response to a slowly moving test charge, using a
second order approximation can then be found [Phys. Lett. A 305, 79 (2002)].
It is also well known [Phys. Plasmas 10, 3484 (2003)] that the dynamical charging
of grains in a dusty plasma enhances the shielding of a test charge. It seems
natural at this stage to seek the combined effects of grain size distribution
and grain charging dynamics to a test charge moving through the dusty plasma.
Here we consider the effects of both grain size distribution and dynamical
grain charging to a test charge moving slowly in a dusty plasma by expressing
the plasma dielectric response as a function of both grain size distribution
and grain charging dynamics. Both analytical as well as the numerical results
are presented. It is interesting to note that the previous results can be
retrieved by choosing appropriate values for different parameters. This kind of
study is relevant for both laboratory and space plasmas.
\end{abstract}


\section{Introduction }

There is currently considerable interest in understanding the physics of
dusty plasmas which in addition to the electrons and ions, also contain
a dust component. The dust grains become charged due to the collection
of ions and electrons from the plasma and can typically acquire
thousands of electron charges (due to high mobility of electrons, dust
grains usually becomes negatively charged). Moreover, the dust charge
does not necessarily remain constant and may continuously fluctuate due
to varying plasma currents that flow onto the dust charge surface. The
currents reaching the dust grain surface depend on the ambient plasma
conditions and the floating potential of the dust particle. In this way
the dust charge becomes a dynamic variable and fluctuates about an
equilibrium charge state. In order to deal with the problem of charging
dynamics, many papers have taken into account this dynamics and
presented their results \cite{1,2,3}, the consequences of
including a dust component have lead to a renewed interest in the
problem of the test charge response. This is important for understanding
the influence of a dust component on the interaction between charged
particles. An important consequence of the potential excited by a moving
charge is the energy loss and braking of the velocity due to the
resultant electric field at the moving charge \cite{4,5}. 

We have investigated the response of a slowly moving test charge in a
dusty plasma in the presence of charging dynamics \cite{6,7} and found that
the dynamical charging of dust grains in a dusty plasma enhances the
shielding of the test charge. The response potential up to the second
order in test charge velocity was found and expressed analytically in
terms of strength functions. A delayed shielding effect due to dynamical
charging was also reported. The linearised dielectric theory was used
and the equilibrium dust distribution was considered to be Maxwellian.
Furthermore, the equilibrium dust particles were assumed to be similar
and all the dust particles were treated as point  particles. But this is
not always true and in general, a size distribution of dust grains is to
be expected both in artificial and natural plasmas  \cite{8,9},
for a Maxwellian distribution and a special class of physically
reasonable size distributions, the dielectric  response function was
shown to be equivalent to that for monosized particles with a
generalised Lorentzian or kappa distribution \cite{10}
Recently, we have taken into account the test charge response of a dusty
plasma with a grain size distribution  \cite{11,12} and have shown that the
form of grain size distribution strongly influences the linear
dielectric response of a test charge in a dusty plasma. The analytical
expressions for the response potential, using a second order
approximation were found and the effects of collisions also
investigated. More recently, A. M. Mirza et al.  \cite{13} extended this work
further and presented analytical as well as numerical results for the
slowing down of a pair of test charge projectiles moving through a
multicomponent dust-contaminated plasma. In their analyses, they found
that the energy loss for the Maxwellian distribution is larger compared
to that for generalised Lorentzian distribution. They also found that
for smaller values of the spectral index $\kappa$, the test charge projectile
gains instead of losing energy. 

In this paper, we have extended our previous work on grain size
distribution  \cite{12} by taking into account the effect of charging dynamics
and present analytical as well as numerical results for the response
potential for a test charge moving through a multicomponent dusty
plasma. 

\section{Plasma Dielectric for a Dusty Plasma}

The linear response of the dusty plasma for an electrostatic disturbance can
be determined through the choice of the plasma dielectric function. Here the
dielectric will include a term for the dynamical charging of the dust grains
and the effect of a specific choice for the size distribution will be taken
into account.

\subsection{Grain Size Distribution}

Here we choose the size distribution $h(a)$ used previously \cite{10}. 
\begin{equation}
h(a) da = h_{0} a^{\beta} \exp \left( - \alpha^3 a^3 \right) da
\label{dist}
\end{equation}
where the constant $h_{0}$ is defined by setting the integrated density to the dust density $n_d$
($h_{0} = 3 n_d \alpha^{\beta+1} / \Gamma((\beta+1)/3) $).
The distribution $h(a)$ has a maximum at $a = a_{0} \equiv (\beta/3)^{1/3} \alpha^{-1}$.
If we let $\alpha \to \infty$ with $\beta \sim \alpha^{3}$ the distribution $h(a)$ tends
to a delta function at $a = a_{0}$ i.e. a monosized distribution with dust grain radius $a_{0}$.
This limit is useful for comparing the general results that will be found here with earlier
results for monosized distributions.
The distribution $h(a)$ can also be transformed to a distribution over grain mass $m$ so that,
\begin{equation}
h(a) da = w(m) dm \equiv w_{0} m^{(\beta-2)/3} \exp \left( - \mu m \right) dm
\label{mass}
\end{equation}
For small sizes $h(a)$ has an approximate power law dependence on the size $a$. A power
law dependence is a simple first approximation if the actual size distribution is not known.
For large masses $w(m)$ is dominated by an exponential decrease with mass (for $\beta=2$
the dependence is purely exponential, as assumed in early work on interstellar dust grains).
These properties motivated this particular choice of size distribution \cite{10}. 
Without charging dynamics this choice for $h(a)$ leads to a dielectric response equivalent
to a kappa distribution \cite{10} with  $\kappa ={{\left( {2\beta +5} \right)} / 6}$.
(For the purely exponential mass dependence, $\beta=2$, the index $\kappa = 3/2$.)

\subsection{Charging Dynamics}

Here we now include charging dynamics with frequencies
$\nu _0 \equiv \Omega_{u0}$ and $\Omega_{v0}$ (defined by Melands\o~ {\sl et al} \cite{15})
that depend on grain size $a$. This leads to a response term with
an integration over grain size with a differential
``charging wavenumber'' $H_{dch}$ given by
$$H_{dch}^2\left( a \right)da\equiv 4\pi {{\Omega_{v0}} 
\over {\Omega_{u0}}}ah\left( a \right)da$$
where $\Omega_{u0}$ and $\Omega_{v0}$ are the frequencies introduced by
Melands\o~ {\sl et al} \cite{15} in the linearized equation for the grain
charge perturbation $q_{d1}$ with a plasma potential perturbation $\phi_{1}$,
\begin{equation}
\frac{\partial q_{d1}}{\partial t}=-\Omega_{u0}q_{d1}
-4\pi\varepsilon_{0}{a}\Omega_{v0}\phi_{1}  \label{dcha}
\end{equation}
In this equation $\Omega_{u0} \equiv \nu_{0}(a)$ is the grain charge
relaxation rate i.e. $\tau_{0} \equiv 1/\nu_{0}$ is the time scale for the grain
charge to come into equilibrium with the undisturbed plasma. 
The total charging wave number  $K_{dch}$ may then be defined by 
$$K_{dch}^2\equiv \int\limits_0^\infty  H_{dch}^2 da$$
For a monosize distribution  $h(a)$ is a delta function and the
above expression reduces to the standard definition  \cite{7}. 
Integrating for $h(a)$ given by the equation~(\ref{dist}) gives
for $H_{dch}$ the expression
\begin{equation}
K_{dch}^2={4 \pi}{ \Omega_{v0} \over \Omega_{u0}} { n_d \over \alpha}
{ \Gamma\left( {{\beta + 2} \over 3}\right) \over
\Gamma\left( {{\beta + 1} \over 3}\right) }
\label{kdch}
\end{equation}

\subsection{Plasma Dielectric}

For a general size
distribution with charge relaxation rate $\nu_0(a)$ that is a function of
the dust grain radius $a$ the plasma dielectric function is,
\begin{eqnarray}
%
%
 & & {
D(K,K\cdot V_t)=1+{{K_{De}^2} \over {K^2}}+{{K_{Di}^2} \over {K^2}} }
\nonumber\\
& & 
%
%
+{{K_D^2} \over {K^2}}\left[ {1+\left( {{{2\kappa } \over {2\kappa -1}}} \right)
\left( {{{\hat K\cdot V_t} \over {V_{td}}}} \right)
Z_\kappa \left( {{{\hat K\cdot V_t} \over {V_{td}}}} \right)} \right]
%
%
%
%
+{1 \over {K^2}}\int\limits_0^\infty  {{{H_{dch}^2\left( a \right)\nu _0\left( a \right)} 
\over {\left( {\nu _0\left( a \right)-iK\cdot V_t} \right)}}}da
\label{a2}
\end{eqnarray}
where $K_D$  and $V_{td}$  are the effective Debye wave-number and effective
thermal velocity for the dust as defined in  \cite{10}. For  $V_t<V_{td}$, the plasma
dispersion function  $Z_\kappa \left( {{{\hat K\cdot V_t} / {V_{td}}}} \right)$  
is given as follows  \cite{16}:
\begin{equation}
Z_\kappa \left( {{{\hat K\cdot V_t} \over {V_{td}}}} \right) =
{{i\sqrt \pi } \over {\kappa ^{3/2}\Gamma \left( {\kappa -{1 \over 2}} \right)}}
\sum\limits_{n=0}^\infty  {\left( {-{1 \over {i\sqrt \kappa }}} \right)^n
{{\Gamma \left( {\kappa +{1 \over 2}\left( {n+2} \right)} \right)} \over 
{\Gamma \left( {{1 \over 2}\left( {n+2} \right)} \right)}}}
\left( {{{\hat K\cdot V_t} \over {V_{td}}}} \right)^n
\label{a3}
\end{equation}
Following the analysis of Melands\o~ {\sl et al} \cite{15}, for a standard model of
the dust charging process, explicit expressions can be found for the frequencies
$\Omega_{u0} \equiv \nu_{0}(a)$ and $\Omega_{v0}$. 
These may be written as \cite{7},
\begin{eqnarray}
\Omega_{v0} &=& \delta_{v0} {a \over \lambda_{De}} \omega_{pi}
\\
\Omega_{u0} &=& \delta_{u0} {a \over \lambda_{De}} \omega_{pi}
\label{nu0}
\end{eqnarray}
where, assuming equal ion and electron temperatures, the numerical constants
are $\delta_{v0} = 2.793$ and $\delta_{u0} = 1.795$. Here we note that, for grain
sizes comparable to the electron Debye length $\lambda_{De}$, these frequencies are
of the order of the ion plasma frequency $\omega_{pi}$.
The frequencies $\Omega_{v0}$ and $\Omega_{u0} \equiv \nu_{0}$ are simply
proportional to the dust size $a$ and the ratio 
$\delta_{0} \equiv \Omega_{v0}/\Omega_{u0} = \delta_{v0}/\delta_{u0} = 1.556$
is independent of the dust size. The last term in equation~(\ref{a2}) can now be
written using these expressions and the size distribution $h(a)$ defined by
equation~(\ref{dist}). There is no obvious simple analytical expression for the
resulting integration, but for a slowly moving test charge the integral can
be expanded as a power series in $V_t$. The individual terms can then be integrated
in terms of the gamma function.

\section{Response to a Moving Test Charge}
For a test charge response in a plasma, the general expression for the
electrostatic potential is given by \cite{14}
\begin{equation}
\phi =\frac{q_{t}}{8\pi ^{3}\varepsilon _{0}}\int \frac{\exp [i{\bf \ K\cdot
r}]}{K^{2}D\left( K,{\bf \ K\cdot V}_{t}\right) } d{\bf K}
\label{a1}
\end{equation}
where $V_{t}$  is the test charge velocity and $D(K,\omega)$ is the plasma
dispersion function. The explicit form of $D(K,\omega)$ depends on the physics
of the dusty plasma. Here $D(K,\omega)$ is chosen to include the effects of
a grain size distribution and charging dynamics.

For a slowly moving test charge 
($V_t<V_{td}$), we can expand the plasma dispersion function (equation~(\ref{a3}))
up to first order and hence rewrite equation~(\ref{a2}) for the dielectric up to
second order in test charge velocity as
\begin{eqnarray}
%
%
%
 & & {
{1 \over {K^2D(K,K\cdot V_t)}} = {1 \over {K^2+K_{eff}^2}}
\left[
{1+iA\left( \beta \right)
{{K_D^2} \over {K^2+K_{eff}^2}}
\left( {{{\hat K\cdot V_t} \over {V_{td}}}} \right)}
\right. }
\nonumber\\
& & 
%
%
\left.  
-B\left( \beta  \right){{K_D^2} \over {K^2+K_{eff}^2}}
{\left( {{{\hat K\cdot V_t} \over {V_{td}}}} \right)^2+
{i} {{\alpha \lambda_{De}} \over {\delta_{u0} \; \omega _{pi}}}{{K_1^2 K} \over {K^2+K_{eff}^2}}
\left( {{{\hat K\cdot V_t}}} \right)
}
\right. 
\nonumber\\
& & 
%
%
\left.
{ 
-C\left( \beta  \right) 
{ \alpha^{2} \lambda_{De}^{2} \over {\delta_{u0}^{2} \; \omega_{pi}^{2}} } 
{{K_1^2 K^{2}} \over 
{K^2+K_{eff}^2}}\left( \hat K\cdot V_t \right)^2}
 + \;
{\cal O}(V_{t}^3) 
\right]^{-1}
\label{a4}
\end{eqnarray}
with the definitions
$$K_1^2={4 \pi \delta_{0}}{ n_d \over {\alpha}},
\ K_{eff}^2=K_{De}^2+K_{Di}^2+K_D^2+K_1^2f\left( \beta  \right),
\ f\left(\beta  \right)=
{{\Gamma \left( {{{\beta +2} \over 3}} \right)} \over 
{\Gamma \left( {{{\beta +1} \over 3}} \right)}}$$
and with
$$
A\left( \beta  \right)=
{{\sqrt \pi } \over {\sqrt {{1 \over 3}\beta +{5 \over 6}}}}
{{\Gamma \left( {{1 \over 3}\beta +{{11} \over 6}} \right)} \over 
{\Gamma \left( {{1 \over 3}\beta +{4 \over 3}} \right)}},
\ B\left( \beta  \right)=
{{4\left( {\beta +4} \right)} \over {2\beta +5}}, \ 
C\left( \beta  \right)=
{{\Gamma \left( {{\beta  \over 3}} \right)} \over 
{\Gamma \left( {{{\beta +1} \over 3}} \right)}}.\ $$
Here the term $K_{1}^2 f(\beta)$ appearing in the definition of $K_{eff}$
is identical to $K_{dch}^2$ given by equation~(\ref{kdch}). The total
charging wave number $K_{dch}$ measures the contribution of charging dynamics
to the total effective shielding wave number $K_{eff}$.
In the above relations $\beta$ is the power law index of the size
distribution for small radii, related to the equivalent kappa
distribution by the relation  $\kappa ={{\left( {2\beta +5} \right)} / 6}$. 
Expanding equation~(\ref{a4}) for the inverse
dielectric function up to second order in test charge velocity  $V_{t}$ and
using in equation (1), we may express the
electrostatic potential as $\phi =\phi _1+\phi _{ch}$ with
\begin{eqnarray}
%
%
 & & {
\phi_1 = {{q_t} \over {8\pi ^3\varepsilon _0}}\int 
{{{\exp [iK\cdot r]} \over {K^2+K_{eff}^2}}}  
\left[ 1-{{iA\left( \beta  \right)K_D^2} \over {K^2+K_{eff}^2}}
\left( {{{\hat K\cdot V_t} \over {V_{td}}}} \right) 
\right. }\nonumber\\
& & \left. 
%
%
+{{B\left( \beta  \right)K_D^2} \over {K^2+K_{eff}^2}}
\left( {{{\hat K\cdot V_t} \over {V_{td}}}} \right)^2
-{{A\left( \beta  \right)^2K_D^4} \over {\left( {K^2+K_{eff}^2} \right)^2}}
\left( {{{\hat K\cdot V_t} \over {V_{td}}}} \right)^2 \right] d{\bf K}
\label{a5}
\end{eqnarray}
and
\begin{eqnarray}
%
 & & {
  \phi_{ch} = {{q_t} \over {8\pi ^3\varepsilon _0}}\int 
{{{\exp [iK\cdot r]} \over {K^2+K_{eff}^2}}}
\left[ -{i}{ {\alpha \lambda_{De}} \over {\delta_{u0} \; \omega_{pi}}  } { {K_1^2 K} \over {K^2+K_{eff}^2}}
\left( {{{\hat K\cdot V_t} }} \right)
\right. }\nonumber\\
& & \left. 
%
%
+ C\left( \beta  \right)
{\alpha^{2} \lambda_{De}^{2} \over {\delta_{u0}^{2} \; \omega_{pi}^{2}} }
{ {K_1^2 K^2} \over {K^2+K_{eff}^2} }
\left( {{ {\hat K\cdot V_t}  }} \right)^2
- { {\alpha^{2} \lambda_{De}^{2}} \over {\delta_{u0}^{2} \; \omega_{pi}^2} } 
{ {K_1^4 K^2} \over {\left( {K^2+K_{eff}^2} \right)^2} }
\left( { {\hat K\cdot V_t} } \right)^2
\right. \nonumber\\
& & \left. 
%
%
- 2 A\left( \beta  \right)
{ {\alpha \lambda_{De}} \over {\delta_{u0} \; \omega_{pi}} }
{K \over V_{td}}
{ {K_D^2K_1^2} \over {\left( {K^2+K_{eff}^2} \right)^2} }
\left( {\hat K\cdot V_t} \right)^2 
\right] d{\bf K}
\label{a6}
\end{eqnarray}
It is to be noted that $\phi_1$ is the same as we found earlier  \cite{12} except for the
definition of $K_{eff}$  which now includes the effect from charging dynamics in
terms of  $K_{1}$, while  $\phi_{ch}$ is the contribution which comes explicitly from
the dust charging dynamics.
%
%
The reader is referred to \cite{12} for the results of equation (\ref{a5}) for ${%
\phi _{1}}$, while in the following we shall present the results for ${\phi
_{ch}}$. The above equation (\ref{a6}) can be written in terms of strength
functions as
%
%
%
\begin{equation}
\phi_{ch}(r,\lambda) = \frac{q_{t}}{8\pi^{3}\varepsilon_{0}}
\left[ V_{t} \, g_{11}(r) \cos \lambda
+ V_{t}^{2} \, \left( g_{20}(r) + g_{22}(r) \cos^{2}\lambda \right) 
+ \;
{\cal O}(V_{t}^3) \right]
\label{response}
\end{equation}
where $\lambda $ is the angle between the test particle velocity 
$\mathbf{V}_{t}$ and the radial vector $\mathbf{r}$. 
The strength functions $g_{ij}(r)$ are given by the following expressions
%
%
\begin{eqnarray*}
g_{11}\left( r\right)  &=&{\pi ^{2}} \frac{{\alpha \lambda _{De}}}
{{\delta _{u0}\;\omega _{pi}}} {K_{1}^{2}} {\exp }\left( -rK_{eff}\right)  \\
%
%
g_{20}\left( r\right)  &=& {\pi^{2} \over 4 K_{eff}} \frac{\alpha^{2}\lambda_{De}^{2}}
{{\delta_{u0}^{2}\;\omega _{pi}^{2}}}
{{K_{1}^{4}} }
\left[ 
 {4 K_{eff}^2 C\left( \beta \right) \over K_{1}^{2}}
 {1 \over r K_{eff}}
 -1
\right]
{\exp}\left(-rK_{eff}\right)
\\
&& + { A\left( \beta \right) \over 2 {V_{td}} } \;
{\frac{{\alpha \lambda_{De}} }{{\delta _{u0}\;\omega _{pi}}}} \;
\frac{{K_{D}^{2}K_{1}^{2}}}{r^{3}K_{eff}^{5}}
\left[ 
 {r K_{eff}} \left( 3 + r^{2}K_{eff}^{2} \right) \Phi
-\left( 3+2r^{2}K_{eff}^{2}\right) \Psi
+6rK_{eff}
\right] \\
%
%
g_{22}\left( r\right)  &=& - {\pi^{2} \over 4 K_{eff}}  \frac{\alpha^{2}\lambda_{De}^{2}}
{{\delta_{u0}^{2}\;\omega _{pi}^{2}}}
{{K_{1}^{4}} }
\left[ 
 {4 K_{eff}^2 C\left( \beta \right) \over K_{1}^{2}}
 {\left( 1 +  r K_{eff} \right) \over r K_{eff}}
  - r K_{eff}
\right]
{\exp}\left(-rK_{eff}\right)
\\
&& - { A\left( \beta \right) \over 2 {V_{td}} } \;
{\frac{{\alpha \lambda_{De}} }{{\delta _{u0}\;\omega _{pi}}}} \;
\frac{{K_{D}^{2}K_{1}^{2}}}{r^{3}K_{eff}^{5}}
\left[ 
rK_{eff}\left( 9+2r^{2}K_{eff}^{2}\right) \Phi  \right. \\
&& \left.
-\left(r^{2}K_{eff}^{2}+3+rK_{eff}\right) \left( r^{2}K_{eff}^{2}+3-rK_{eff}\right)
\Psi   +2rK_{eff}\left( 9+r^{2}K_{eff}^{2}\right) 
\right]
\end{eqnarray*}
%
%
%
where the following relations defining $\Phi (r K_{eff})$ and $\Psi (r K_{eff})$
 in terms of exponential integrals \cite{7} have been introduced,
\begin{equation}
\Phi \left( r K_{eff}\right) =\exp \left( r K_{eff}\right) E_{1}\left(
r K_{eff}\right) -\exp \left( -r K_{eff}\right) E_{i}\left( r K_{eff}\right) 
\end{equation}
\begin{equation}
\Psi \left( r K_{eff}\right) =\exp \left( r K_{eff}\right) E_{1}\left(
r K_{eff}\right) +\exp \left( -r K_{eff}\right) E_{i}\left( r K_{eff}\right) 
\end{equation}
$\Phi (y)$ and $\Psi (y)$ (for $y>0$) are directly defined as principal
parts of integrals (here, for real $y$, equivalent to taking the real part)
as follows,
\begin{equation}
\Phi \left( y\right) =Re\left\{ -\int_{0}^{\infty }\frac{2\,t\,\exp (-yt)\,dt}
{1-t^{2}}\right\} 
\end{equation}
\begin{equation}
\Psi \left( y\right) =Re\left\{ \int_{0}^{\infty }\frac{2\,\exp (-yt)\,dt}
{1-t^{2}}\right\} 
\end{equation}
From these definitions it follows that $\Phi(y) = d \Psi(y) / dy$ and that
$\Psi(y) = d \Phi(y) / dy +2/y$.
The functions $\Phi (y)$ and $\Psi (y)$ introduced here are closely related
to the auxiliary functions $f(y)$ and $g(y)$ used in the analysis of the
Sine and Cosine Integrals \cite{a22}. As for $f(y)$ and $g(y)$ asymptotic
forms may be found for $\Phi (y)$ and $\Psi (y)$, 
\begin{equation}
-\frac{1}{2}\,\Phi \left( y\right) \sim {y}^{-2}+3!\,{y}^{-4}+5!\,{y}%
^{-6}+7!\,{y}^{-8}+O\left( {y}^{-10}\right) 
\end{equation}
\begin{equation}
\frac{1}{2}\,\Psi \left( y\right) \sim {y}^{-1}+2!\,{y}^{-3}+4!\,{y}%
^{-5}+6!\,{y}^{-7}+O\left( {y}^{-8}\right) 
\end{equation}
%
%
\section{Discussion}

In equations~(\ref{a4}) and (\ref{a6}) the combination of terms 
$\tau \equiv {\alpha \lambda_{De}} / {\delta_{u0} \; \omega_{pi}}$ is equal to
$1/\nu_0(\alpha^{-1})$ (from equation~(\ref{nu0}) where $\Omega_{u0} \equiv \nu_0$).
Therefore $\tau$ is the relaxation time for the charge on a dust grain with radius
$\alpha^{-1}$ to reach equilibrium with the ambient plasma, and $V_{t}\tau$ is the
distance travelled by the test charge in this time. As remarked above,
if we let $\alpha \to \infty$ with $\beta \sim \alpha^{3}$ the distribution $h(a)$ tends
to a monosized distribution with dust grain radius $a_{0} \equiv (\beta/3)^{1/3} \alpha^{-1}$.
Putting $\tau \equiv (\beta/3)^{1/3} \tau_{0}$, the test charge response $\phi_{ch}$
given by equation~(\ref{response}) may be shown to reduce to the known results for a
monosized distribution with a charge relaxation time $\tau_{0}$ \cite{6,7}.

%
%

\section{Acknowledgement}

The authors would like to thank their colleagues at the Alfv\'{e}n
laboratory for useful discussions and suggestions. This work was partially
supported by the Swedish Research Council.

%
%

\end{document}